\documentclass{article}

\usepackage{arxiv}

\usepackage[utf8]{inputenc} 
\usepackage[T1]{fontenc}    
\usepackage{hyperref}       
\usepackage{url}            
\usepackage{booktabs}       
\usepackage{amsfonts}       
\usepackage{nicefrac}       
\usepackage{microtype}      
\usepackage{graphicx}
\usepackage{textcomp}
\usepackage{xcolor}
\usepackage{cite}
\usepackage{caption}
\usepackage{pgfpages}
\pgfpagesuselayout{resize to}[a4paper]
\def\BibTeX{{\rm B\kern-.05em{\sc i\kern-.025em b}\kern-.08em
    T\kern-.1667em\lower.7ex\hbox{E}\kern-.125emX}}
    
\title{Enabling Technologies for 6G Future Wireless Communications: Opportunities and Challenges}

\author{
  Samar Elmeadawy \\
   \And
 Raed M. Shubair \\
}

\begin{document}
\maketitle


\section{Introduction}
\begin{figure}[h]
\centering
\includegraphics[scale=0.4]{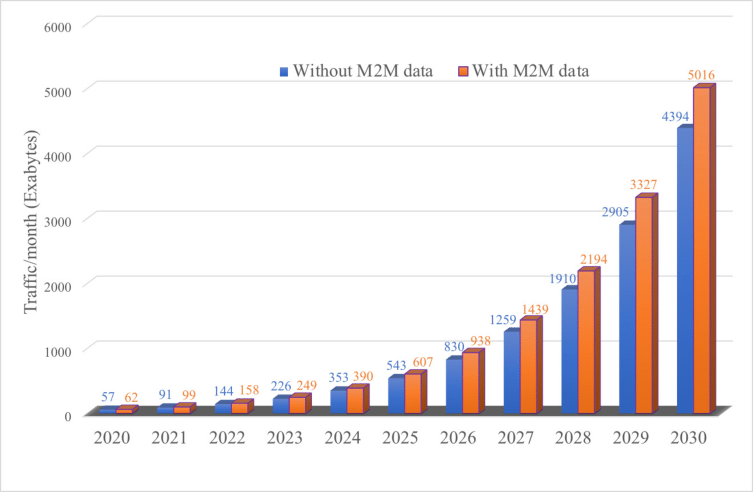}
\caption{ITU Global mobile data traffic Prediction \cite{tariq2019speculative}}
\label{statistics}
\end{figure}

Almost every ten years, a new communication system has been introduced, improving the QoS, providing new features and introducing new technologies. 
Although 5G is not officially launched yet, researchers have turned their attention to 6G communication system. The reason is that 5G provides a high standard infrastructure enabling a variety of technologies such as; self-driving cars, AI, mobile broadband communication, IoT and smart cities. However, the usage of smart devices is increasingly growing each year and the data traffic usage will be exponentially increasing as in Fig. ~\ref{statistics}, which puts constraints on the 5G communication network.

These constraints open the door for a new communication system providing more capacity, extremely low latency, high data transmission, secure error-free communication and full wireless coverage. Table ~\ref{comparison}
compares the main specifications and technologies in both 5G and 6G.
6G will be able to connect everything, integrate different technologies and applications, support holographic, haptic, space and underwater communications and it will also support the Internet of everything, Internet of Nano-Things and Internet of Bodies \cite{zhang20196g}.

\begin{table}[h]
    \centering
    \caption{Comparison between 5G and 6G}
    \label{comparison}
    \bgroup
    \def\arraystretch{1.4}
    \begin{tabular}{|c||c|c|}
        \hline
        \textbf{Characteristic} & \textbf{5G} & \textbf{6G} \\
         \hline
         \hline
         \textbf{Operating frequency} & 3 - 300 GHz & upto 1 THz \\
         \hline
         \textbf{Uplink data rate} & 10 Gbps & 1 Tbps \\
         \hline
         \textbf{Downlink data rate} & 20 Gbps & 1 Tbps \\
         \hline
         \textbf{Spectral efficiency} & 10 bps/Hz/$m^2$ & 1000 bps/Hz/$m^2$ \\
         \hline
         \textbf{Reliability} & $10^-5$ & $10^-9$ \\
         \hline
         \textbf{Maximum mobility} & 500 km/h & 1000 km/hr \\
         \hline
         \textbf{U-plane latency} & 0.5 msec & 0.1 msec \\
         \hline
         \textbf{C-plane latency} & 10 msec & 1 msec \\
         \hline
         \textbf{Processing delay} & 100 ns & 10 ns \\
         \hline
         \textbf{Traffic capacity} & 10 Mbps/$m^2$ & 1 - 10 Gbps/$m^2$ \\
         \hline
         \textbf{Localization precision} & 10 cm on 2D & 1 cm on 3D \\
         \hline
         \textbf{Uniform user experience} & 50 Mbps 2D & 10 Gbps 3D \\
         \hline
         \textbf{Time buffer} & not real-time & real-time \\
         \hline
         \textbf{Center of gravity} & user & service \\
         \hline
         \textbf{Satellite integration} & No & Fully \\
         \hline
         \textbf{AI integration} & Partially & Fully \\
         \hline
         \textbf{XR integration} & Partially & Fully \\
         \hline
         \textbf{Haptic communication integration} & Partially & Fully \\
         \hline
         \textbf{Automation integration} & Partially & Fully \\
         \hline
    \end{tabular}
    \egroup
\end{table}

Future 6G wireless communication systems will benefit from the knowledge gained and solutions achieved from several technologies, methods, and applications \cite{elayan_wireless_2017,khan_ultra-compact_2017,elayan_multi-layer_2017,khan_pattern_2017,ibrahim_compact_2017,elayan2017wireless,nwalozie2013simple,al2005direction,shubair2005robust,belhoul2003modelling,al2005computationally,shubair2004robust,shubair1993closed,al2003performance,elayan2018terahertz,al2003investigation,khan_pattern_2016,elayan2017photothermal,el2016design,elayan2017bio,elayan2018end}.  In this report, some emerging technologies and applications introduced and developed by the 6G communication technology are presented in section ~\ref{technologies} and the main challenges facing the achievement of the 6G goals are addressed in section ~\ref{challenges}.

\clearpage

\section{Emerging technologies and Applications}
\label{technologies}
\begin{figure}[h]
\centerline{\includegraphics[scale=0.34]{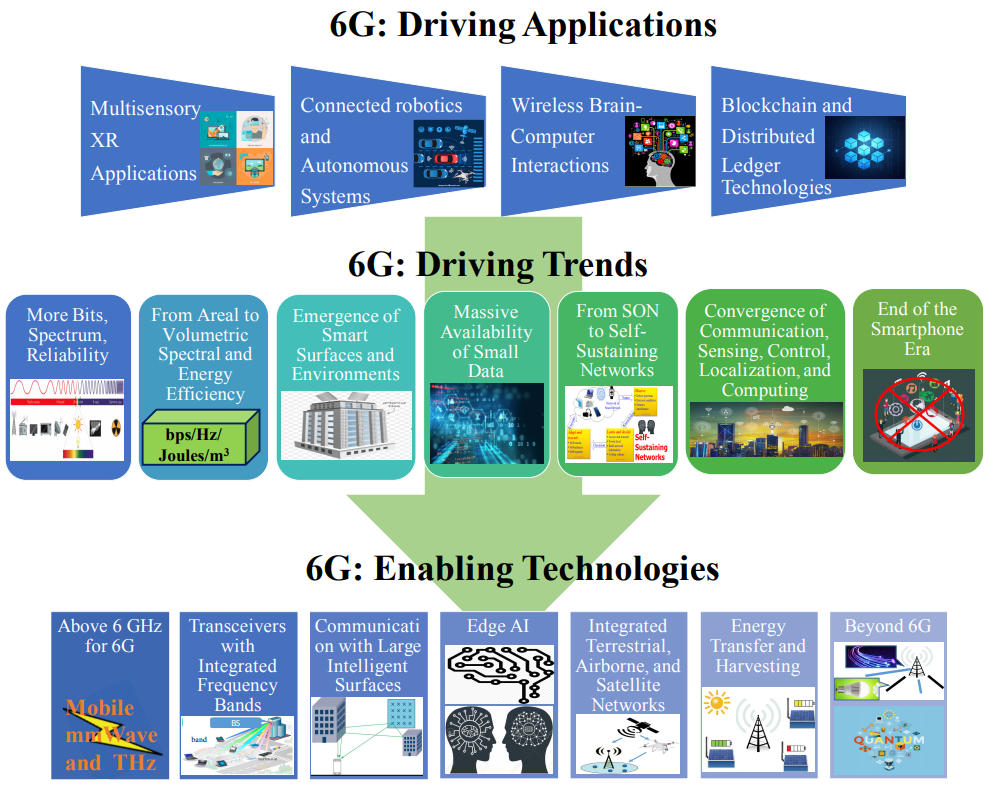}}
\caption{6G main Applications, trends, and technologies \cite{saad2019vision}}
\label{6Gvision}
\end{figure}

Every communication system opens the door to new features and applications. 5G was the first generation to introduce AI, automation and smart cities. However, these technologies were partially integrated. 6G is introducing more technologies and applications providing higher data rates, high reliability, low latency and secure efficient transmission. Fig. ~\ref{6Gvision} shows the main applications, trends and technologies introduced in 6G. In this section, some of these technologies and applications 6G are discussed.

\subsection{TeraHertz Communication}
The RF band is almost full and it is not able to support the increasingly growing demand in wireless communications technology. The THz band, ranging from 0.1 THz to 10 THz, will play a crucial role in 6G supplying more bandwidth, more capacity, ultra-high data rates and secure transmission. The THz band will support the development of minuscule cells in nanometer to micrometre dimensions supplying very high-speed communications within a coverage area of up to 10 m \cite{akyildiz2014terahertz} and supporting the Internet of Nano-things \cite{elayan2018terahertz,elayan2016graphene}. Technologies using frequency bands below 0.1 THz can not support Tbps links, therefore, 6G will be the first wireless communication system supporting Tbps for high-speed communication.

\subsection{Cell-Free Communication}
\label{cellfreecomm}
Unmanned Aerial Vehicles (UAV) were proposed to be used in other generations in places where there is no infrastructure. However, this technology will be fully used in 6G allowing cell-free communication. When the user equipment (UE) moves from one cell coverage to another, the user‘s call should be transferred to the other cell. This handover might be unsuccessful and in some occurrences, the user‘s call is terminated and the QoS will be reduced in the system. 6G will end the problem of cell coverage as the UE will be connected to the whole network, not a specific cell. Using UAV will allow integrating different technologies allowing the UE to utilize the technology having the best coverage without any manual configurations on the device \cite{giordani2019towards}.

\subsection{Artificial Intelligence}
Artificial Intelligence (AI) was not involved in 4G or any previous generations. It is partially supported by 5G making difference in the telecommunications world opening the doors for emerging remarkable applications such as \cite{alhajri2018classification,alhajri2016classification,alhajri2019indoor,alhajri2018machine,UCIML24GHZ}. However, AI will be fully supported in 6G for automation. It will be involved in the handover, network selection and resources allocation improving the performance, especially in delay-sensitive applications. AI and machine learning are the most important technologies in 6G \cite{strinati20196g}.

\subsection{Holographic Beamforming (HBF)}
\begin{figure}[htbp]
\centering
\centerline{\includegraphics[scale=0.36]{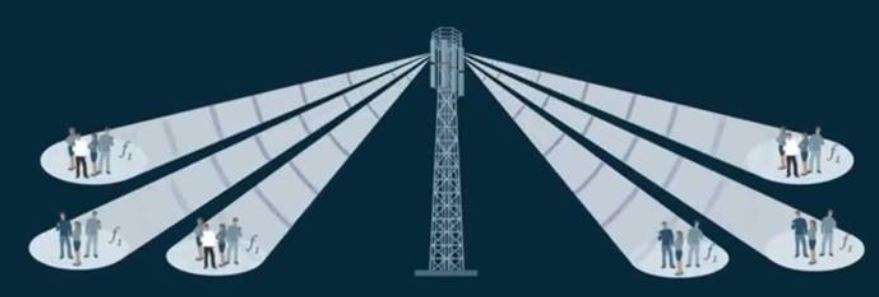}}
\caption{Holographic Beamforming  \cite{b1}}
\label{holographic}
\end{figure}

Beamforming is using a directed narrow beam with a high gain for transmitting and receiving using antenna arrays by focusing the power in a minimized angular range. It offers better coverage and throughput, higher signal to interference and noise ratio (SINR) and it could be used to track users. Holographic beamforming is an advanced beamforming approach utilizing Software-Defined Antenna (SDA).
Holographic refers to using a hologram to achieve beam steering by the antenna, where the antenna is like a holographic plate in an optical hologram; RF signals from a radio flow into the back of the antenna and scatter across its front, where tiny elements adjust the shape and direction of the beam as in Fig. ~\ref{holographic}. SDAs are cheaper, smaller in size, lighter and require less power compared to the traditional phased arrays or MIMO systems \cite{black2016holographic}. As C-SWaP (Cost, Size, Weight and Power) are considered as the main challenges in any communication system designs, using SDAs in HBF will allow flexible and efficient transmitting and receiving in 6G.

\subsection{Extended Reality}
Extended reality (XR) is a new umbrella term including Virtual Reality (VR), Augmented Reality (AR) and Mixed Reality (MR). 
VR is a computer-simulated reality experience using a headset that generates sounds and images creating an imaginary world.
AR uses the real world and adds to it using a specific device such as the mobile phone. Audios, videos, Global Positioning System (GPS) could be used to create an interactive environment. Pokémon is a well-known example of AR. 
MR merges between the real and the virtual worlds creating a complex environment. XR is all the real and virtual environment combined. 6G will be very useful for this feature due to the strong connectivity, high data rate, high resolution and low latency \cite{piran2019learning}.

\subsection{Blockchain technology}
The data in the blockchain technology are represented as distributed blocks connected to each other and cryptographically secured. Blockchain will be used in managing and organizing big data and in managing huge connectivity in 6G. It will be used also in spectrum sharing allowing the users to share the same spectrum solving the problem of huge spectrum requirements in 6G and guaranteeing secure, low cost, smart and efficient spectrum utilization. Integrating the blockchain with AI and using Deep reinforcement learning \cite{dai2019blockchain} will improve the QoS allowing smart-resources sharing, implementing an advanced caching scheme and making the network more flexible.

\subsection{Automation}
Currently, researchers focus on automation, robotics and autonomous systems. 6G will support these technologies providing direct communication between them and the server and direct communication between them, i.e.: robot to robot communication and robot to the server communication.
Full automation will be provided by 6G including automatic control processes, automatic systems and automatic devices.
6G will support the existence of Unmanned Aerial Vehicles (UAV) \cite{li2018uav} which will be used in wireless communications providing high data rates instead of the traditional base stations (BS).

\subsection{Wireless Power Transfer}
Wireless energy transfer will be involved in 6G, providing suitable power to the batteries in devices such as; smartphones and sensors \cite{jung2015qoe}. The base stations in 6G will be used for transferring power as Wireless Information and Energy Transfer (WIET) uses the same fields and waves used in communication systems. WIET is an innovative technology that will allow the development of batteryless smart devices, charging wireless networks and saving the battery life-time of other devices. 

\subsection{Wireless Brain-Computer Interface}
Recently wearable devices are increasingly used, some of them are brain-computer interface (BCI) applications. BCI applications involve smart wearable headsets, smart embedded devices and smart body implants \cite{saad2019vision}. Using BCI technologies, the brain will easily communicate with external discrete devices which will be responsible for analyzing brain signals and translating them. BCI also will involve affective computing technologies, in which devices will function differently depending on the user’s mood. BCI applications were limited because they require more spectrum resources, high bit rate, very low latency and high reliability. However, 6G will support more applications such as the five sense information transfer, in which 6G will transfer the data generated by the five senses of the human allowing the interaction with the environment.

\subsection{Healthcare}
\begin{figure}[htbp]
\centering
\centerline{\includegraphics[scale=0.36]{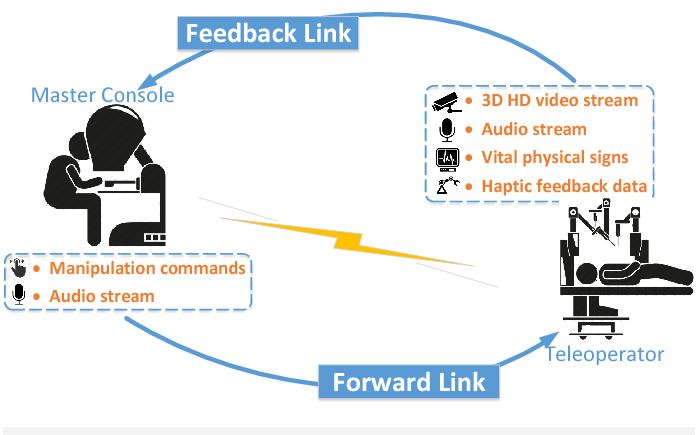}}
\caption{A loop explaining the communication between the master console and the tele-operator \cite{zhang2018towards}}
\label{surgery}
\end{figure}

The lack of electronic healthcare in other wireless communication technology was because of low data rate and time delay. 6G will provide secure communication, high performance, ultra-low latency, high data rate and high reliability enabling the full existence of remote surgeries as in Fig. ~\ref{surgery} through XR, robotics, automation and AI \cite{giordani2019towards}. Also, the small wavelength due to the THz band supports the communication and the development of nanosensors allowing developing new nanosized devices to operate inside the human body \cite{elayan2017terahertz,elayan2017photothermal,elayan2018end,shubair2015vivo,elayan2017multi}.
\clearpage
\section{Research Challenges}
\label{challenges}
To fulfil the world demands in technology, some challenging requirements in 6G wireless communication should be achieved. In this section, the main challenging topics are investigated and discussed.

\subsection{TeraHertz Band}
\label{THz}
The main challenge in the 6G wireless communication system is the THz band. Although it provides high data rates, the high frequencies make overcoming the high path loss a great issue. For long-distance communications, the atmospheric absorption and propagation loss are very high. This is an important issue that needs to be addressed. 
Because of the broad bandwidth, new multipath channel models need to be developed to overcome the problem of frequency dispersion \cite{akyildiz2014terahertz}. The existing modulation and coding techniques are not sufficient for the THz band. Therefore, implementing new modulation and coding techniques is challenging research. 
New transceivers should be designed to operate on the high-frequency band supporting the very large bandwidth, high power, high sensitivity and low noise figure to overcome the atmospheric losses. Also, health and safety issues due to high power and frequency are great challenges facing the researchers.

\subsection{Device Capabilities}
Devices were not compatible with all the wireless communication technologies. Recently, companies are working on devices supporting 5G, these devices should be able to support 6G and all the different wireless communication generations as discussed in section ~\ref{cellfreecomm}. Devices supporting 6G wireless communication technology should tolerate the 1 Tbps data rate and the high operating frequencies. Devices should also support interacting with different devices using Device to Device communication, AI and XR \cite{chowdhury20196g}. 
In these days smartphones are almost being charged daily utilizing more energy than before. Billions of devices other than smartphones will be connected to the 6G network, therefore, efficient energy transferring methods should be considered especially wireless energy transfer methods \cite{pon2019printed}, and devices connected should be developed so that they can support different charging methods. These capabilities in the devices are costly and challenging.

\subsection{Network Security}
6G wireless communications network will connect not only smartphones but also smart devices used in automation, AI, XR, smart cities and satellites. The security approaches used in 5G will not be sufficient in 6G, and hence new security techniques with innovative cryptographic methods should be considered including the physical layer security techniques and integrated network security techniques \cite{yang20196g} with low cost, low complexity and very high security.

\subsection{Transceiver and Antenna Designs}
For each wireless communication technology, there is a specific transceiver and antenna designs supporting the specifications of that technology as mentioned in section ~\ref{THz}. it was a challenge in 5G to develop devices of millimetre components. However, in 6G it will be more challenging. 6G wireless communication technology supports high-frequency band in THz and supports spectrum and resources sharing. The transceivers should be able to support these technology having the antenna designed with the required size; nanometer to micrometre components satisfying the holographic beamforming requirements as in \cite{goian2015fast,alhajri2015hybrid,shubair2016new,alhajri2018accurate}. Metasurface-based transceivers \cite{tang2019wireless} could be the solution to this issue to increase the throughput and the QoS. However, integrating metasurface with OFDM-MIMO is a great challenge.
\clearpage
\section{Conclusion}
5G telecommunication technology that will be launched in 2020 will not fulfil the increasingly growing demands in 2030. Therefore researches in 6G should be conducted to be able to reach its goals by 2030. In this report, new features in 6G and the possible applications and technologies that will be deployed in 6G are provided. Main challenges in the technologies in 6G are presented. It is concluded 6G will improve the network performance, integrate different technologies and increase the QoS providing super-smart society with everything connected to the network.

\bibliographystyle{IEEEtran}
\bibliography{references}

\end{document}